\documentclass[10pt, letterpaper, a4paper]{amsart}
\usepackage{amsmath}
\usepackage{latexsym}
\usepackage[all]{xy}
\usepackage{color}

\include{amslatex}

\usepackage[margin=1.4in]{geometry}

\numberwithin{equation}{section}
\begin{document}
\begin{title}[Spacetimes of maximal proper acceleration and muon mean lifetime]
{Spacetimes of maximal proper acceleration and muon mean lifetime}
\end{title}
\date{\today}
\maketitle
\begin{center}
\author{Ricardo Gallego Torrom\'e\footnote{Email: rigato39@gmail.com}}
\end{center}

\begin{abstract}
In this paper, it is shown that in spacetimes of maximal proper acceleration, there is a reduction of the mean lifetime of the muon. We consider several candidates for the maximal proper acceleration. For all of them, the induced corrections are small, but for the case when the maximal proper acceleration is identified with the Schwinger's acceleration, the effect is detectable in future muon collider facilities.
\end{abstract}
\bigskip

{\bf Keywords}: Maximal proper acceleration;  Muon lifetime.
\section{Introduction}
First proposed by E. Caianiello in the context of a geometric approach to quantum mechanics \cite{Caianiello, Caianiello 1992},  the conjecture on the existence of a maximal proper acceleration has emerged in different contexts \cite{Ricardo Nicolini 2018}.
It arises, for instance, in several gravitational theories and quantum gravity frameworks \cite{Brandt1983,Parentani Potting,Frolov Sanchez 1991,Rovelli Vidotto, Ricardo05b, Ricardo06, Ricardo2014, Ricardo 2024b, Harikumar et al. 2020,Harikumar et al. 2021, Harikumar et al. 2022} and in quantum electrodynamics in the form of the acceleration scale associate with the Schwinger limit \cite{Sauter, Schwinger}. It is also consequence of the general principles of quantum mechanics and special relativity \cite{Caianiello 1984}, and also arise in several classical theories of electrodynamics \cite{Caldirola,Ricardo2015,Ricardo 2017,Ricardo 2024,Ricardo 2025}. Such an ubiquity of maximal proper accelerations, as well as the fact that it is a Lorentz invariant concept, provides to the conjecture of physical relevance.

The maximal proper acceleration scale associated with gravity is extremely large compared with current acceleration scales achievable in laboratories. The case is different for the maximal proper acceleration scales associated with certain models of electrodynamics or quantum theoretical arguments. Indeed, consequences of maximal proper acceleration in high order jet electrodynamics,  potentially observable in current accelerator and laser-plasma accelerator facilities, have been discussed in references \cite{Ricardo 2019,Ricardo 2024,Ricardo 2025, Ricardo Milan 2025}. However, such consequences rely on additional modelling assumptions. In this context, the present paper highlights a direct implication of maximal proper acceleration which is on the boundary of being testable with current or planned accelerator facilities. In particular, the effect of maximal proper acceleration in the reduction of the muon mean lifetime. Muon systems constitute a very valuable tool in the search of deviations from standard spacetime geometry. Indeed, the study of the muon mean lifetime leads to accessible tests of modified dispersion relations \cite{Lobo Pfeifer} and former studies leaded to maximal acceleration corrections to the Lamb shift of muonic hydrogen, as it was discussed in the context of Caianiello's maximal acceleration geometric theory in \cite{Chen et al.}.

The present work is developed in the framework of spacetimes of maximal proper acceleration. This is a classical geometry theory, in the sense that there is no superpositions of spacetime metrics and events are associated with points of a four dimensional spacetime manifold. Although the theory of spacetimes of maximal proper acceleration was initially motivated by the problem of the non-general covariance of Caianiello's geometry \cite{Caianiello et al. 1990}, soon developed independently as a theory of generalized high order jet geometry and high order jet fields
\cite{Ricardo2007, Ricardo2015, Ricardo 2019, Ricardo 2020, Ricardo 2024, Ricardo 2025, Ricardo 2017}. However, the concept of spacetime of maximal proper acceleration has a broader scope of applicability than high order jet electrodynamics, since it can embrace theories containing a notion of maximal proper acceleration consistent with the premises of a classical spacetime geometry. In this paper, after introducing the notion of spacetime of maximal proper acceleration, we apply the theory to several candidates for a maximal proper acceleration. Besides high order jet electrodynamics, we consider the geometries associated with Caianiello's geometry and the possibility that the maximal proper acceleration is identified with Schwinger's acceleration. In this last case, the reduction of the muon life time is potentially observable in future muon collider facilities.

\section{Spacetimes of maximal proper acceleration}
In a spacetime of maximal proper acceleration, the proper time functional depends on the second jet of the test particle used to probe the spacetime. Events are well described by points of a four-dimensional smooth manifolds $M_4$ and test particles have a well-defined, smooth world line in $M_4$. Although the theory is classical in the sense that superposition of spacetimes or quantum of gravity do not appear, fields and geometry acquire a larger {\it plasticity} than in relativistic theories. If $\vartheta:I\to M_4$ is a smooth curve, then $\,^2\vartheta(t)=\,\left(t,\vartheta^\mu(t),\frac{d{\vartheta}^\mu(t)}{dt},\frac{d^2{\vartheta}^\mu(t)}{dt^2}\right)$ determines a point of the second jet manifold $J^2M_4$. Note that the parameter $t$ one of the coordinates of $\,^2 \vartheta (t)$ and also the parameter of the curve $\vartheta:I\to M_4$. The proper acceleration is defined by the expression $a^2:=\,\eta(\nabla_{\dot{\vartheta}}\dot{\vartheta},\nabla_{\dot{\vartheta}}\dot{\vartheta}),$ where the metric $\eta$ is the limit when $A_{\textrm{max}}\to +\infty$ of the metric of maximal acceleration $g$.
 $\nabla$ is the covariant derivative operator associated with its Levi-Civita connection.
Fixing the parameter of the curve by the proper parameter $\tau$ associated to the metric $\eta$, the metric structure of a spacetime of maximal proper acceleration is of the form
\begin{align}
g|_{\,^2\vartheta(\tau)} :=\,\left(1-\, \frac{\eta|_{\vartheta(\tau)}(\nabla_{\dot{\vartheta}(\tau)}\dot{\vartheta}(\tau),
\nabla_{\dot{\vartheta}(\tau)}\dot{\vartheta}(\tau)
)}{A^2 _{\textrm{max}}}\right)\eta|_{\vartheta(\tau)}
\label{maximalaccelerationmetric}
\end{align}
 It determines the physical proper time along $\vartheta$, which is given by the expression
\begin{align}
s [\vartheta] :=\,\int_{\vartheta} \,d\tau\,\left( -g|_{\,^2\vartheta(\tau)}(\dot{\vartheta},\dot{\vartheta})\right)^{1/2} = \,\int_{\vartheta} \,d\tau\,\left(1-\frac{a^2}{A^2_{\textrm{max}}}\right)^{1/2},
\label{propertime g}
\end{align}
where the value of the metric of maximal proper acceleration is $A_{\textrm{max}}$, in the sense that $a^2 <\,A^2_{\textrm{max}}$ is the sufficient condition for $\eta$ and $g$ being causally equivalent along $^2\vartheta$.

\section{Scales of maximal proper acceleration}
The value of the maximal proper acceleration depends on the type of interaction involved and on the theory in question. In the case of gravity, the maximal acceleration scale is of order $A_{\textrm{max}}\sim \,\left(c^7/\hbar\,G\right)^{1/2}\sim 10^{52}\, m/s^2$, very far from any experiment feasible in current facilities \cite{Brandt1983,Parentani Potting,Frolov Sanchez 1991,Rovelli Vidotto}.

Closer to the current experimentally accessible scales is the maximal proper acceleration in high order jet electrodynamics \cite{Ricardo2012, Ricardo 2017}. In such a theory, the value of the maximal acceleration is
\begin{align}
 A^2_{\textrm{max}}\leq \,\left(\frac{3}{2}\,\frac{m_0}{q^2}\right)^2 ,
\label{valueofthemaximalacceleration}
\end{align}
where $m_0$ is the inertial mass of the particle and $q$ its electric charge. One is conveyed to identify the value of the proper acceleration with the right hand side, although this is not strictly proved by the theory.
This value of the acceleration coincides, except by a numerical factor of $1/2$, with the value found by Caldirola in his theory of the electron \cite{Caldirola} and with the value found by Goto et al. in their analysis of the motion of a charged particle in a gravitational field \cite{Goto et al. 2010}. For an electron, this maximal acceleration is $A_{\textrm{max}}(e^-)\sim 10^{31}\, m/s^2$, while for a muon, it is of order $A_{\textrm{max}}(\mu^-) \sim 10^{34}\, m/s^2$.

Phenomenological consequences of the maximal proper acceleration for electron bunches in particle accelerators considered as a sole particle have been discussed in \cite{Ricardo 2019, Ricardo 2024, Ricardo 2025}. This was a consequence of the modified maximal proper acceleration for a bunch, that from the relation \eqref{valueofthemaximalacceleration} reduces to $A^2_{\textrm{max}}(N)=\frac{1}{N}\,\left(\frac{3}{2}\,\frac{m_0}{q^2}\right)^2$. However, such reduction of the maximal proper acceleration is based on additional assumptions of the model of the bunch. In the present work we observe that the maximal proper acceleration for an individual muon leads to a reduction of the mean lifetime respect to the value provided by relativistic quantum field theory. This effect does not depend upon additional modelling considerations of the muons bunch structure, except on the assumption that a perturbative treatment on the variable $\epsilon =\,a^2/A^2_{\textrm{max}}$  can be used at the level where the geometric theory is applicable.

Let us consider two more theories where a maximal proper acceleration arises.
It was discussed by Caianiello, as a result of an geometric approach to quantum mechanics, that a particle with mass $m$ at rest should show a maximal proper acceleration of the form \cite{Caianiello,Caianiello 1984}
\begin{align}
A_{C}\leq\,2\,\frac{mc^3}{\hbar}.
\label{Maximal acceleration of Caianiello 2}
\end{align}
The theory goes further and identifies the maximal proper acceleration with the right hand side of the inequality.
For an electron, we have that $A_{C}[e^-]\sim \,10^{29}\,m/s^2$, while for a muon it is order $A_{C}[\mu^-]\sim\,10^{31}\,m/s^2$.

Finally, let us consider the acceleration scale associated with Schwinger limit in quantum electrodynamics. In this case, the value of the maximal acceleration for a muon in an external Schwinger field, by means of applying Lorentz force equation as a first approximation, is given by the expression
\begin{align}
A_{S}[\mu]=\,\frac{m_e}{m_\mu}\,\frac{ \,m_e\,c^3}{\hbar}.
\label{Schwinger acceleration for the muon}
\end{align}
The argument leading to the relation \eqref{Schwinger acceleration for the muon} justifies the equality, rather than an inequality. Note that the Schwinger's acceleration has its origin in purely quantum effects and hence, its imposition as the value of the maximal proper acceleration in the spacetime metric structure must be understood as an effective model, valid for accelerations far from such a non-linear, quantum regime. Thus accepting this heuristic argument and considering Schwinger's acceleration as an effective maximal proper acceleration that can be plug-in the metric of maximal proper acceleration, due to the factor $m_e/m_\mu$, Schwinger's acceleration for the muon $A_S[\mu^-]$ becomes more {\it accessible} to experimental probe than the corresponding Caianiello's maximal acceleration $A_C[m_\mu]$. Indeed, one has $A_{S}[\mu]\sim 10^{27}\,m/s^2$,
four order of magnitude smaller than Caianiello's maximal acceleration for the muon, and seven orders of magnitude smaller than the maximal acceleration of the high order jet electrodynamics.
\section{Reduction of the muon mean lifetime due to a maximal proper acceleration}
In the perturbative regime where $\epsilon:=\,a^2/A^2_{\textrm{max}}\ll 1$, it is natural to assume that the muon mean lifetime is given by the inverse of the transition rate $\Gamma (\mu^-)$. In this regime, the transition rate  $\Gamma (\mu^-)$ has a perturbative form in $\epsilon$, treated as a parameter. Thus, we can write
\begin{align*}
{\Gamma} (\mu^-)=\, \Gamma_{rel} (\mu^-)+\, \Gamma_1\,\frac{a^2}{A^2_{\textrm{max}}}+\mathcal{O}(\frac{a^2}{A^2_{\textrm{max}}})^2,
\end{align*}
where $\Gamma_{rel}(\mu^-)$ is the relativistic rate of the transition and $\Gamma_1$ is a constant. Let us assume that the relativistic muon mean lifetime $1/\Gamma_{rel} (\mu)$ of a muon before decaying can be casted as the proper time of a point particle along a specific world line $\bar{\vartheta}:I\to M_4$,
\begin{align}
1/\Gamma_{rel} (\mu^-)=\,\int_{\bar{\vartheta}} \eta(\vartheta' (t), \vartheta'(t))^{1/2}\, dt =\,\int_{\bar{\vartheta}} \,d\tau=\,\tau_{rel} [\bar{\vartheta}].
\end{align}
Thus, the muon mean lifetime in a spacetime of maximal proper acceleration in the perturbative regime where $a^2/A^2_{\textrm{max}}\ll 1$ is of the form
\begin{align}
1/\Gamma (\mu^-) =\,\int_{\bar{\vartheta}} \,d\tau\,\left(1-\frac{a^2}{A^2_{\textrm{max}}}\right)^{1/2},
\end{align}
with no further perturbative contributions.
If the proper acceleration $a$ is constant, then
\begin{align*}
1/\Gamma (\mu^-) =\,\left(1-\frac{a^2}{A^2_{\textrm{max}}}\right)^{1/2} \,\int_{\bar{\vartheta}} \,d\tau =\,\left(1-\frac{a^2}{A^2_{\textrm{max}}}\right)^{1/2} \,1/\Gamma_{rel} (\mu^-).
\end{align*}
Therefore, the existence of a maximal proper acceleration reduces the muon mean lifetime with respect to the expected relativistic value. The same consequence holds good for any unstable particle that can be described kinematically by a smooth world line.
 In the perturbative domain, the reduction of the muon lifetime  is of the form
\begin{align*}
\Delta (\mu^-):=\,\frac{1/\Gamma (\mu^-)-1/\Gamma_{rel} (\mu^-)}{1/\Gamma (\mu^-)}\approx -\frac{1}{2}\,\frac{a^2}{A^2_{\textrm{max}}}+\mathcal{O}\left(\left(\frac{a^2}{A^2_{\textrm{max}}}\right)^2\right)
\end{align*}

Experimentally, the value of the muon mean lifetime is
$1/\Gamma (\mu^-)=\,2.1969811(22)\times 10^{-6} \,s .$, that shows an accuracy of order $10^{-8}$. Therefore,
in order to have observable signatures of maximal proper acceleration in the lifetime of the muon, it is necessary that the maximal proper acceleration leads to corrections at the scale $\frac{1}{2}\,\frac{a^2}{A^2_{\textrm{max}}}\approx 10^{-8}$ or larger with respect to the evaluation of the muon lifetime using relativistic quantum electrodynamics or any other relativistic field theory.

In order to illustrate the magnitude of the effect,consider first the case of high order jet electrodynamics. Since the maximal proper acceleration for a muon is of order $A_{\textrm{max}}[\mu^-]\sim 8\times \,10^{34} \,m/s^2$,  the proper acceleration must be of order $a\sim 8.8\, \times 10^{30}\, m/s^2$ or larger to reach an observable correction with the current accuracy of the muon lifetime. Such acceleration scales are far from current and planned facilities. For Caianiello's maximal proper acceleration, the scale of the acceleration required to test the required size of the correction to the muon lifetime is of order $10^{27}\,m/s^2$. This is still far from current experimental access.
However, for the case of Schwinger acceleration $A_{S}[\mu^-]\sim \,10^{27}\,m/s^2$, the required experimental scale of acceleration is of order $a\sim\,10^{23}$ to $10^{24}\,m/s^2$.

Let us consider a circular accelerator operating at $10^4\, GeV$. These energy scales are planned in future muon colliders. In this case, the acceleration will be of order $a\sim \,10^{27}/R[m] ms^{-2}$, where the curvature radius $R[m]$ is given in meters. In such a situation, the proper acceleration could be of order $a\sim \,10^{23}$ for $R[m]=\,10^4$. Hence, if the maximal proper acceleration is associated with the Schwinger limit,
this leads to a reduction in the muon mean lifetime of the order $10^{-8}$, which is on the range to be observable.
\section{Discussion}
In this paper, it has been shown that the existence of a maximal proper acceleration implies a reduction of the muon mean lifetime.
The effect in question is discussed in the framework of spacetimes of maximal proper acceleration \cite{Ricardo2015, Ricardo 2024, Ricardo 2025}. This is a classical theory, since it applies to accelerations far from the quantum dynamical regime and to smooth spacetime world lines. On the other hand, the quantum regime, in the form of the Schwinger limit, poses a limit acceleration scale of order $10^{27} \,m/s^2$ beyond which the geometric theory is not applicable. Also not the compatibility of the bounds \eqref{maximalaccelerationmetric} and \eqref{Maximal acceleration of Caianiello 2} with the Schwinger maximal proper acceleration \eqref{Schwinger acceleration for the muon}.

There are several candidates for the maximal proper acceleration.
Besides high order jet electrodynamics, we have considered two more candidates for the value of the maximal proper acceleration. The first has been Caianiello's maximal proper acceleration \cite{Caianiello,Caianiello 1992,Caianiello 1984}. In this case, the theory leads to corrections of the muon lifetime which are too small to be observed with current or planned facilities.
The case is different if one adopts Schwinger's acceleration as the maximal proper acceleration. In this case, if the reduction in the muon life time is of order $10^{-8}$, then the scales of proper acceleration needed will be reachable in projected accelerator facilities.

 The above discussion assumes that the determination of the muon mean lifetime is of order $10^{-8}$. If an increase in the experimental accuracy on the determination of the muon mean lifetime is achieved, then the effect described in this paper should be visible at lower energies and it will make accessible to test high order jet electrodynamics and Caianiello's theory of quantum geometry.

The theory assumes the existence of a spacetime world line where the muon  mean lifetime is its proper time parameter. For the case of the maximal proper acceleration in high order jet electrodynamics, this is a natural assumption. However, for such a theory, the corrections to the muon mean life are too small to be observable with current technologies. On the other hand, although considering classical geometries, we have also applied the geometric framework of spacetimes of maximal proper acceleration to theories where the maximal proper acceleration has a quantum origin. This is legitimated as long as the value of the proper acceleration $a$ is small compared with the {\it quantum threshold scale}.

 The notion of spacetime of maximal proper acceleration is an effective notion where spacetime appears as the prism through which we discuss phenomena. The effective value of the maximal proper acceleration depends on the kind of phenomena and the interactions involved. Of course, one can consider the supreme of all the possible maximal proper accelerations and then one will have a definition of spacetime of maximal proper acceleration that does not depend upon the particular phenomena, such a supreme probably being related with the maximal proper acceleration associated with gravity. We think, however, that this is not the way the concept is useful in the present context.

 The theory considered in this paper has to address the following dialectic issue. From one side, increasing the scale of the acceleration implies that the effect caused by the maximal proper acceleration is larger and larger. From the other side, there is a limit of applicability of the theory due to quantum effects. Indeed, the maximal acceleration scale associated with the classical maximal proper acceleration is larger than the associated with the quantum maximal proper acceleration of Caianiello and Schwinger theories.
 The limit imposed by the quantum theory is even stronger, since the notion of spacetime acceleration is not well defined, at least for the use made in the current paper, for an individual quantum system. A high order jet geometry spacetime metric is defined on the second jet $J^2M_4$, which explicitly makes use of the covariant, geometric definition of proper acceleration. Therefore, according to conventional interpretations of quantum mechanics, the theory loses its applicability when the scale of accelerations is close to scales where quantum dynamics deviates from classical spacetime dynamics. Such a scale is the Schwinger's acceleration scale for the discussed phenomena.

 Now we look at this issue from other perspective. Schwinger's acceleration scale and Caianiell0's maximal proper acceleration offer the possibility to test the notion of acceleration at the deep quantum domain and in particular, to test alternative interpretations of quantum phenomena such that the notion of trajectory of an individual system can be applied at such scales. This is the case of certain {\it emergent quantum mechanic schemes}, for instance, for Bohm-de Broglie theory \cite{Bohm}, Nelson's stochastic mechanics \cite{Nelson Dynamical Theories} and Hamilton-Randers theory \cite{Ricardo05b,Ricardo06,Ricardo2014}. According to Bohm-de Broglie theory, an individual quantum system has associated a classical, smooth spacetime world line. Hence, the notion of acceleration is properly defined. In Nelson's theory, the individual system has associated a continuous, non-smooth world line. Hence the standard definition of acceleration fails at the spacetime level, but one can try to define alternative definitions of spacetime acceleration through {\it balance conditions}. In contrast, in Hamilton-Randers theory, although there is still an spacetime world line associated to the individual quantum system, it has a larger degree of discontinuity that accounts of the discontinuous quantum jumps phenomenology \cite{Ricardo2014}. In Hamilton-Randers theory, one cannot expect to have a classical notion of spacetime proper acceleration for a quantum system. Therefore, we observe that in the dynamical regimen of acceleration reaching the quantum limits, the theory of maximal proper spacetime and its applications on that regime, specifically, the effect on the muon mean life time, leads to a theoretical guide and tool to test the phenomenology of emergent quantum mechanics frameworks.

 \subsection*{Acknowledgements} The author would like to thank professors V. Petrillo and L. Serafini for calling m\emph{}y attention on the acceleration scales reached in standard storage rings.
\small{
}

\end{document}